\documentclass[prd,superscriptaddress,twocolumn,floatfix,nofootinbib]{revtex4}

\usepackage{natbib} 
\usepackage{graphicx}
\usepackage{bm}
\usepackage{epsfig}
 
\def\dvk{\frac{{\rm d}^3k}{(2\pi)^3}}

\def\dvx{{\rm d}^3x}

\def\x{\bm{x}}
\def\k{\bm{k}}

\def\vd{\bm {d}}

\def\kt{\bm{k}_{\perp}}

\def\kp{{k}_{\parallel}}

\def\Mpc{\rm Mpc}

\newcommand{\beq}{\begin{equation}}
\newcommand{\eeq}{\end{equation}}
\newcommand{\bal}{\begin{aligned}}
\newcommand{\eal}{\end{aligned}}
\newcommand{\beqa}{\begin{eqnarray}}
\newcommand{\eeqa}{\end{eqnarray}}

\begin{document}

\title{Dark Energy from Large-Scale Structure Lensing Information
}

\author{Tingting Lu}
\affiliation{Department of Astronomy and Astrophysics, University of Toronto, M5S
3H4, Canada}
\affiliation{Canadian Institute for Theoretical Astrophysics, University of
Toronto, M5S 3H8, Canada}
\author{Ue-Li Pen}
\affiliation{Canadian Institute for Theoretical Astrophysics, University of
Toronto, M5S 3H8, Canada}
\author{Olivier Dor\'e}
\affiliation{Canadian Institute for Theoretical Astrophysics, University of
Toronto, M5S 3H8, Canada}


\date{\today}



\label{firstpage}

\begin{abstract}	

Wide area Large-Scale Structure (LSS) surveys are planning to map a
substantial fraction of the visible universe to quantify dark energy
through Baryon Acoustic Oscillations (BAO).  At increasing redshift,
for example that probed by proposed 21-cm intensity mapping surveys,
gravitational lensing potentially limits the fidelity (Hui et al.,
2007) because it distorts the apparent matter distribution.  In this
paper we show that these distortions can be reconstructed, and actually
used to map the distribution of intervening dark matter.  The lensing
information for sources at $z=1-3$ allows accurate reconstruction of
the gravitational potential on large scales, $l \lesssim 100$, which is
well matched for Integrated Sachs-Wolfe (ISW) effect measurements of
dark energy and its sound speed, and a strong constraint for modified
gravity models of dark energy.

We built an optimal quadratic lensing estimator for non-Gaussian sources,
which is necessary for LSS.  The phenomenon of ``information saturation''
(Rimes \& Hamilton, 2005) saturates reconstruction at mildly non-linear
scales, where  the linear source power spectrum $\Delta^2\sim 0.2-0.5$. We
find that steeper power spectra, saturate
more quickly.  We compute the effective number densities of independent
lensing sources for LSS lensing, and find that they increase rapidly
with redshifts. For LSS/21-cm sources at $z\sim 2-4$,  the lensing
reconstruction is limited by cosmic variance at $l\lesssim 100$.

\end{abstract}
 

\maketitle

\section{Introduction} \label{INTRO}

The standard cosmological model has achieved substantial quantitative
success through precision cosmology.  It requires a mysterious and
dominant dark energy component, which is not physically understood.
Or it may be a hint that Einstein's General Relativity is not the correct
theory to describe our universe.

New quantitative tests are needed to improve our understanding of the nature
of dark energy or modified gravity.  In this paper we show how
high-z Baryon Acoustic Oscillations (BAO) surveys, in addition to measuring kinematic cosmology, can
also constrain the dynamics of perturbations through weak gravitational lensing.
A direct measurement of the gravitational potentials can determine
the sound speed of the dark energy field \citep{2004PhRvD..69h3503B}.

Recently, modified gravity models have been studied quantitatively.
In order to pass local precision gravity tests, and match the Cosmic
Microwave Background (CMB) fluctuations, these models tend to differ in their
predictions only on very large scales and at low redshifts.  

Direct all-sky lensing maps are observationally challenging,
since the signals are weak, and systematics are difficult to control.
However, quadratic estimators of CMB maps have recently demonstrated
its feasibility on very large scales \citep{2007PhRvD..76d3510S}.
Similar techniques have been proposed for lensing of diffuse
structures\citep{2004NewA....9..417P}, and will be explored in further
detail in this paper.

The kinematic history of the universe is being mapped out using the
standard metre stick of BAO.  A range
of surveys are underway and have been proposed to produce a coarse
image of large-scale structure (LSS) over a substantial sky area and
redshift range.  While these surveys are not directly suitable for lensing
mapping, the maps they produce will be distorted by gravitational lensing
\citep{2007PhRvD..76j3502H}.  Since the intrinsic correlations is known
to be statistically isotropic, the lensing induced changes can be used
to reconstruct the dark matter map, in analogy with lensing of the CMB.
The CMB is a source screen with well known lensing properties, for which
lensing and ISW results in a non-zero three point function.  As we show
in this paper, the kernels are not perfectly matched, and the dynamical
dark energy properties can be better measured with sources at a range
of redshifts, as is the case for 21cm LSS.  This general approach is
known as lensing tomography.

Some of the proposed fast BAO surveys \citep{2008PhRvL.100i1303C, 2008MNRAS.383.1195W,2008arXiv0805.4414T} are based on an
intensity mapping approach, where the LSS is mapped without detection
of individual galaxies over a large fraction of the sky.  For such
surveys, point source based lensing mappings are not applicable \citep{2005PhRvL..95x1302Z}, and the LSS lensing is the only available tool. Lensing reconstruction using Gaussian random fields as sources
is now well understood for both 2-D and 3-D structures \citep{2004NewA....9..417P,2004NewA....9..173C,
2006ApJ...653..922Z, 2007MNRAS.381..447M,2008MNRAS.388.1819L,
2008arXiv0801.2571B}. \citet{2008arXiv0801.2571B} have explored the possibility of
applying 21-cm lensing from low and high redshifts to the dark energy
constraint. While their method is based on the Gaussian assumption, it
was shown that the neglect of non-Gaussianity could generate orders of magnitude difference
in the noise estimation \citep{2008MNRAS.388.1819L} (hereafter LP08). 

In general, lensing reconstruction is a quadratic function of the density field,
which is quantified by its power spectrum. 
\citet{2005MNRAS.360L..82R}
(hereafter RH05) showed that the Fisher information available in the power
spectrum saturates at trans-linear scales, and stays several orders of magnitude below the
Gaussian Fisher information content. In this paper, we combine the
information from these approaches, and construct the optimal quadratic
lensing estimators for non-linear sources fields from N-body simulations.

We find that the reconstruction noise of lensing from the simulated
sources decreases with reduced experimental noise at linear scales,
saturates at quasi-linear scales, and drops again at highly non-linear
scales.  There is a plateau region at the trans-linear regime, which is
analogous to what was found with the Fisher information content in the initial
amplitude of matter power spectrum by
\citet{2005MNRAS.360L..82R}. Both effects are caused by the
non-Gaussianity (or non-linearity) introduced by gravitational clustering during structure
formation. We found the saturation scale is a steadily increasing function
of the shape of the linear power spectrum.

This non-Gaussianity makes the 21-cm lensing less promising than it
first appears.  However, because of the abundance of 21-cm sources, the Signal-to-Noise ratio
(S/N) of 21-cm lensing at high redshift $z\sim 1-6$ is still competitive compared
with other surveys. The effective number density of sources defined by
the independent number of cells in the 21-cm source, increases quickly
with redshift.  At redshift $4-6$, this number density could be $9 \,
{\rm arcmin}^{-2}$.

The paper is organized as follows: We overview the progress in the
LSS  information and lensing studies in section~\ref{LSS}. The optimal lensing estimator is 
introduced in section~\ref{LENS}. The numerical methods are presented in 
section~\ref{NUM}. 
The results are discussed in section~\ref{DISC}. We conclude in section~\ref{CONC}. 

\section{Information and lensing from the LSS} \label{LSS}

The lensing of Gaussian random fields can be described in several
ways. For the purposes of this paper, we are in a regime where the
fluctuation scale of the sources is smaller than the structures of 
interest in the lenses.  In this limit, lensing has two effects on an
image: convergence ($\kappa$) and shear ($\gamma$).

As we will find through detailed calculations, these two quantities
carry a comparable amount of information, dependent on the slope of
the power spectrum of sources.  For pedagogical purposes, we first
consider the convergence, that is the stretching of the image 
changes the amplitude of fluctuations at a fixed apparent angular scale. By
comparing the power spectrum in different patches, we can measure the relative convergence.

The accuracy of this procedure is proportionate to the precision
with which the power spectrum can be measured.  This problem has been
studied in terms of Fisher Information by RH05, and their results directly
translate into the accuracy of lensing reconstruction.  After a technical
calculation, we find that the effective number of independent lensing
sources is equal to the RH05 Fisher Information, scaled by a coefficient
of order unity, which depends on the slope of the power spectrum.

A related problem is the impact of lensing on the sky averaged power spectrum
\citep{2007PhRvD..76j3502H,2008PhRvD..77f3526H,2008PhRvD..77b3512L}.
There had been concern that the lensing at low redshift may impact the ability to accurately
measure the matter power spectrum at high redshift.  To first order, there is an equal
number of positive and negative convergence patches, so the net effect is
second order. Since we can measure the first order effect on each patch,
we can actually correct its impact, and cancel off the second order bias.

\section{Lensing reconstruction} \label{LENS}

In the lensing of diffuse background, what is observed is the distorted brightness 
temperature distribution. As discussed in LP08, all lensing estimators are quadratic terms of temperature fields. 
The lensing field (convergence $\kappa$, shear $\gamma$, or 
deflection angle $\vd$) can be reconstructed with the product of two smoothed 
temperature fields.  For different estimators, the window functions could also be 
scalars or vectors. 

How to choose the smoothed window is just a matter of optimization.
There are also progressive levels of optimization to be considered: Is the source 
isotropic? Is the window function isotropic? For Gaussian distributed temperature field, 
the optimal window function can be solved analytically.  \citet{2002ApJ...574..566H} 
formulated the optimal quadratic deflection angle estimator (OQDE) for 2-D CMB lensing, 
and later \citet{2006ApJ...653..922Z} generalized it to 3-D 21-cm lensing. The OQDE 
surprisingly has the same form at all scales, even though an additional scale dependent 
normalization factor needs to be applied at the last step.  
In an alternative approach, by optimizing the reconstruction noise in the
zero-mode, we derived  the  optimal Gaussian window for the
convergence and shear fields separately in LP08. 

In reality, 21-cm sources are not Gaussian fields. 
For simplicity, we 
use the 3-D dark matter distribution  to
represent the 21-cm emission at the same redshift. We investigated both OQDE and our estimators by numerical tests with simulated 
non-Gaussian source fields and mocked Gaussian ones. For Gaussian sources, we found the combined result of our two estimators has exactly 
the same S/N as the OQDE on all scales. This agrees with our expectation because of the scale independence 
of the optimal window functions. If one window function is optimal at one scale, it is also 
optimal at other scales.  For non-Gaussian sources, however, the optimal window function derived
from Gaussian assumption is not optimal any more. The lensing reconstruction noise
could be underestimated by orders of magnitude if we use the results
from Gaussian sources, which could be calculated analytically by
applying Wick's theorem. By choosing appropriate weights, our combined
estimator has better performance than  the OQDE. 

In this paper, we will further explore the optimization of  non-Gaussian sources.
We will develop the optimal non-Gaussian estimator,  which can only be constructed by 
numerical measurements from a large sample of simulations. The
reconstruction noise of optimal lensing estimators for non-Gaussian
21-cm sources are  closely related to  the Fisher information contained in non-linear matter power spectrum described 
by RH05. They indicated that the Fisher information for the initial 
amplitude of matter power spectrum could be written as the sum of the inverse of covariance 
matrix of power spectrum estimates, multiplied by the partial derivative of power 
spectrum with respect to the initial amplitude. They have revealed the
cumulative Fisher information results up to a maximum scale, which enters the calculation
as the upper limit scale of the covariance matrix.  They found a very interesting phenomenon: 
the Fisher information increases at linear scales, and then the growth becomes very slow at the 
quasi-linear region (which they label as information plateau), and the Fisher information starts
to rise again in highly non-linear scales. 

In this context, Fisher information means how accurately the amplitude
of the power spectrum can be measured. As we  will discuss later in section \ref{NGOQE},  lensing can be detected by the change 
of the source power spectrum.  We will also prove that the optimal
window function of non-Gaussian sources actually contains the inverse covariance 
matrix of power
spectra.  Thus the reconstruction noise of lensing roughly estimates
how much information is contained in the source power spectrum: the lower
the noise level achieved,  the more information is
gained. Instrumental noises of 21-cm experiments  can be
approximately treated as hard cut-offs at some scales, then we could inspect the reconstruction noises with different cut-off scales.
When the sources are Gaussian, the covariance matrix is diagonal and could be expressed as 
a square term of the source power spectrum, and the optimal window function reverts to the form 
we found in LP08. Our numerical results with Gaussian estimators had
similar behavior as RH05: the reconstruction S/N increased first and
reached a peak at trans-linear scales. At non-linear scales, the S/N dropped again.   
We will explain later, the decline of S/N is an artificial effect by using the non-optimal window
function for non-Gaussian sources. In this paper, we will show that the optimal
non-Gaussian estimator also leads to a plateau in trans-linear scales in the S/N plot, and like 
Fisher information of matter power spectrum amplitude, the  curve falls again in non-linear scales.

The covariance matrix of non-linear power spectrum is measured from an ensemble of N-body
simulations. RH05  showed that the covariance of 
power spectrum can only be measured with at least hundreds of independent simulations.  Therefore 
we generated around 100 or more different sources, depending on the redshift, by running the same 
number of N-body simulations.

\subsection{Optimal non-Gaussian estimator} \label{NGOQE}

As we explained in LP08, for the lensing of a diffuse background,
all unbiased estimators are two-point functions of the source temperature field 
with each points convolved with a window function, 
no matter which quantity is reconstructed: the convergence, shear, deflection angle or 
the potential. 
\beq
E(\x) \equiv \tilde T_{w1}(\x) \tilde T_{w2}(\x) \,,
\eeq
where $\tilde T_{w1}, \tilde T_{w2}$ are two convolved temperature fields by some window functions, e.g., 
\beq
\tilde T_{w1}(\x)=\int {\rm d}^2x' \tilde T(\x') W_1(\x-\x') \,.
\eeq
Here $\tilde T$ is the lensed temperature field, and we use $T$ to represent the intrinsic
unlensed temperature field.
We will keep using this convention throughout the paper: the variables with tilde mean 
the quantities with lensing, while those without tilde mean the original ones.
To optimize the estimator, one minimizes its variance
\beq
\sigma^2(E) = \langle{\tilde T_{w1}^2 \tilde T_{w2}^2}\rangle \,,
\eeq
which is a four-point function of the temperature distribution. Except
for a Gaussian distribution, four-point functions are in general not
analytical, and can not be calculated from two-point statistics.  

In LP08, we derived optimal estimators of the 
convergence and shear for a Gaussian source distribution. 
Since the optimal window function for Gaussian sources (hereafter optimal
Gaussian estimator or optimal Gaussian window function) does not
depend on the scale, we were able to look for its form in the
limit of a slowly varying $\kappa$ (or $\gamma$). We then applied a
maximum likelihood method to solve the optimal window. The calculation
was done analytically assuming that the covariance matrix of the power
spectrum is diagonal and can be obtained from the power spectrum,
which is the implication of Wick's theorem. The variance of 
optimal Gaussian estimators rises when the resolution of the
observation is improved,  which shows the optimal Gaussian estimators
are far from optimal for a non-Gaussian source distribution.


To find the optimal estimator for non-Gaussian sources, we pursue
the calculation in a slightly difference approach. We adopt a minimum
variance method, demonstrate it with the Gaussian case and generalize
it to the non-Gaussian case. First, we solve the optimal estimator
for a constant $\kappa$ and a Gaussian source, i.e.,
$\kappa(\x)=\kappa_0$. We then generalize the estimator to the optimal
non-Gaussian estimator of varying $\kappa$ and $\gamma$.

The lensed power spectrum from observations can be Taylor expanded in
first order as:  
\beqa
\tilde P^{\rm tot}(\k) &=& \tilde P(\k) +P^N(\k) \nonumber \\
                       &\approx& P(\k) + P^N(\k) + 
                            \left . {\partial \tilde P(\k) \over \partial \kappa} 
                            \right |_{\kappa=0} \kappa \,,
\label{eq: Pexpand}
\eeqa
where the noise $P^N=P^e+P^s$ includes both the instrumental noise, $P^e$, and the statistical 
fluctuation of the 
power spectrum, $P^s$, which is the sample variance for a Gaussian source.
In LP08, we have defined the derivative of the lensed power spectrum with respect
to $\kappa$, i.e.,  $G(\kappa, \k)\equiv{\partial \tilde P(\k) \over \partial \kappa}$. For convenience, we will use $G$ to represent its value at $\kappa=0$, therefore
\beq
G(\k)  \approx  2P(\k)+\Delta P(\k) \,,
\eeq
where $\Delta P=P'k({k_{\perp}^2/k^2})$,
$P'(k)={\rm d} P(k)/ {\rm d} k$, $k_{\perp}$ means the component of $k$ in the transverse
plane of the line of sight, as defined in LP08. 
Note that here $P(\k)$ is the expectation value of source power spectrum estimate, and 
$\tilde P(\k)$ is the lensed power spectrum of one realization.  In
the absence of lensing, the power spectrum in one realization is
measured by an estimator $\hat P(\k)$, which is equal to the sum of $P(\k)$ and $P^N(\k)$. 
We frequently use the approximation that 
$\langle{(\tilde P^{\rm tot}(\k)-P(\k))^2}\rangle\approx \langle{(\hat P(\k)-P(\k))^2}\rangle$ 
in the reconstruction noise estimation.  
To simplify the algebra, we first consider the case $P^e=0$, and
$P^N=P^s$. It will later be straightforward to add the instrumental noise.

\subsubsection{Discrete case}

Following the convention of
\citet{1992LNP...408...65B}, we use subscript $[...]_{\k}$ to describe discrete quantities, and parenthesis $[...](\k)$ to 
describe continuous quantities. For example, 
\beqa
& &\tilde T(\k) =\int _{-\infty}^{\infty} {\rm d}^3 x \tilde T(\x) e^{-{\rm i} \k \cdot \x}  \,, \nonumber \\
& & \tilde T_{\k} = \sum_j \tilde T(\x_j)e^{-{\rm i} \k \cdot \x_j}  \,.
\eeqa
Note that the observed temperature field is from the contribution of lensed brightness temperature field and noise: $\tilde T=\tilde T_b +n$.  
The expectation value of the source power spectrum is $P_{\k}^{}=\langle{T_{b\,\k}^{} T^{\star}_{b\,\k} }\rangle$, and
$T_{b \,\k}$ is the discrete Fourier transform of $T_b(\x)$. 
If the source brightness temperature follows a Gaussian distribution, the estimators of the
unlensed power spectrum $\hat P_{\k}$ at different $\k$ are independent, and
their variance can be calculated as 
${\rm Var}(\hat P_{\k})=\langle{{P^N_{\k}}^2\rangle= \langle{P^s_{\k}}^2}\rangle 
=P_{\k}^2/n_{\k_b}$, where $n_{\k_b}$ is the number of modes in the $\k$ frequency bin. 

We first consider the discrete case with an independent set of Fourier
frequencies $\k$. For every frequency, an estimator of $\kappa$ can be constructed as:
\beqa
\hat \kappa_{\k} &=& \frac{\tilde P^{\rm tot}_{\k} -P_{\k} -P^N_{\k} }{G_{\k}} \nonumber \\
                 &=& \kappa_0 - n_{\k} \,,
\eeqa
where $n_{\k}=P^N_{\k}/G_{\k}$. Note $\hat \kappa_{\k}$ 
is not the Fourier transform of $\hat \kappa(\x)$,
which has a constant value, but the measurement of $\kappa_0$ at frequency
$\k$. The optimal estimator of $\kappa_0$ should be the total
contribution from all $\k$ appropriately weighted. 
The minimum variance estimator of $\kappa_0$ is
\beq 
\hat \kappa_0= {\sum \hat \kappa_{\k}/\sigma^2_{\k} \over \sum 1/\sigma^2_{\k}} \,,
\eeq
where $\sigma^2_{\k}=\langle{n_{\k}n_{\k}^{\star}}\rangle= P^2_{\k}/ n_{\k_b}G_{\k}^2$. 
Therefore, the weight at each $\k$ is inverse proportional to the $\sigma^2_{\k}$.
The reconstruction noise of the minimum variance estimator is now
\beq
{\rm Var}(\hat \kappa_0)= {1\over \sum 1/\sigma^2_{\k}} \,.
\eeq

More generally, in the non-Gaussian case, power spectrum at different $\k$ are correlated. 
We can write all variables in the form of matrices and vectors:
\beq
{\bf G} \kappa_0 = \tilde {\bf P}^{\rm tot}-{\bf P}-{\bf P}^N \,,
\eeq
where $\bf G$, $\bf P$ and ${\bf P}^N$ are the $N_{\k}\times 1$ matrix composed by 
$G_{\k}$,$P_{\k}$ and $P^N_{\k}$ respectively, and $N_{\k}$ is the number of all frequencies.
The minimum variance (least square) estimator can be written as
\beq
{\bf G}^t  {\bf C^{-1}} {\bf G} \hat \kappa_0 = {\bf G}^t {\bf C^{-1}}  
  (\tilde {\bf P}^{\rm tot}-{\bf P} - {\bf P}^N) \,,
\eeq
where ${\bf C}_{\k,\k'} =\langle{P^N_{\k} P^N_{\k'}}\rangle$ is the covariance
matrix for the power spectrum.  
Since the terms introduced by lensing do not dominate in the noise, we
have neglected their contribution in the covariance matrix ${\bf C}$. 
All noise related
calculations in the following text will use $\hat P$ instead of
$\tilde P^{\rm tot}$, while the estimator of lensing quantities will always use
$\tilde P^{\rm tot}$.
Therefore the minimum variance estimator for the non-Gaussian source is
\beq
\hat \kappa_0 = ({\bf G}^t {\bf C}^{-1} {\bf G})^{-1} 
             {\bf G}^t {\bf C}^{-1}  (\tilde {\bf P}^{\rm tot}-{\bf P}-{\bf P}^N)  \,.
\label{eq:mve}
\eeq
and the variance of the estimator is
\beq
{\rm Var}(\hat \kappa_0) = ( {\bf G}^t {\bf C}^{-1} {\bf G} )^{-1} \,.
\eeq

\subsubsection{Continuous case}

We now consider the continuous case. Assuming the source is a
cube with physical length $L$ in each dimension,
by definition $\tilde T(\k)=\tilde T_{\k}L^3$,  $P(\k)=P_{\k}L^3, G(\k)=G_{\k}L^3$ and 
${\bf C}(\k,\k')={\bf C}_{\k,\k'}L^6$. 
As such, taking the continuum limit leads to
\beqa
\hat \kappa_0 &=& {1\over Q}   \int \dvk {{\rm d}^3 k'\over (2\pi)^3} {\bf C}^{-1}(\k,\k') G(\k') 
(\tilde P^{\rm tot}-P-P^N)(\k) \nonumber \\
&=& \int \dvk \tilde P^{\rm tot}(\k) {\cal F}(\k) -V \,,
\eeqa
where 
\beqa
{\cal F}(\k)& = &  {1\over Q} \int {{\rm d}^3 k'\over (2\pi)^3} {\bf C}^{-1}(\k,\k') G(\k') \label{eq:w_ng} \,, \\
Q & = &\int \dvk {{\rm d}^3 k'\over (2\pi)^3} {\bf C}^{-1}(\k,\k') G(\k) G(\k') \,, \\
V & = & \int \dvk (P(\k)+P^N(\k)) {\cal F}(\k) \,.
\eeqa
V is the mean variance of the smoothed temperature field. Note that when the source is Gaussian, 
the optimal filter is 
${\cal F}(\k)= G(\k) /P(\k)^2/ Q$. This is the estimator presented in
LP08. Using $\tilde T(\k)\tilde T^{\star}(\k) = (2\pi)^3 \delta^{\rm
  3D}(0)\tilde P^{\rm tot}(\k)=L^3 \tilde P^{\rm tot}(\k)$ and 
Parseval's theorem, where $\delta^{\rm 3D}$ is the 3-D Dirac delta function, 
we can write the calculation in real space instead of Fourier space
\beqa
\hat \kappa_0 
 &= & L^{-3} \int \dvk \tilde T(\k) \tilde T^{\star}(\k) {\cal F}(\k) -V \nonumber\\ 
 &=& L^{-3} \int \dvx \tilde T_{w1}(\x) \tilde T_{w2}(\x) -V \nonumber \\
                     &= & \tilde T_{w1} (\x) \tilde T_{w2} (\x) -V  \,,
\eeqa
where the two window functions to smooth the temperature field are the decomposition of
$\cal F$, i.e., $W_1W_2=\cal F$.
The noise of $\hat \kappa_0$ is
\beq
{\rm Var}(\hat \kappa_0) = L^{-6} \int \dvk \int {{\rm d}^3 k'\over (2\pi)^3} {\bf C}(\k,\k') 
{\cal F}(\k) {\cal F}(\k') \,,
\eeq
where ${\bf C}(\k,\k')=\langle{(\hat P(\k) -P(\k))(\hat P(\k')-P(\k'))}\rangle$ is the 
covariance matrix of the power spectrum. 
Similar to Eq. (\ref{eq:mve}) in the discrete case, we have neglected the contribution from lensed terms in the estimate of the covariance matrix ${\bf C}$.
It is clear that the noise
for $\hat \kappa_0$ equals $Q^{-1}$. ${\bf C}(\k,\k')$ will be measured from simulations.

Since
\beqa
\tilde P(\kt,\kp) 
&=& |\mathbf J|^{-1} P(\mathbf J^{-1}\kt,\kp) \nonumber \\
&\approx& (1+2\kappa) [P(k)+
 \Delta P(\k)(\kappa+\gamma_1 \cos2\theta_{\kt} \nonumber \\
& &\,\,+\gamma_2 \sin2\theta_{\kt}) ] \,,
\label{eq:angular}
\eeqa
where $\theta_{\kt}$ is the angle between $\kt$ and the transverse
coordinate (LP08), the first order Taylor expansion gives
\beqa
G^{\gamma_1}(\k)& = &\Delta P(\k) \cos2\theta_{\kt} \,, \\
G^{\gamma_2}(\k)& = &\Delta P(\k) \sin2\theta_{\kt}\ .
\eeqa
We will rewrite the equation in spherical
coordinates using $\k=(k,\theta,\phi)$, $k_{\perp}^2/k^2=\sin^2{\theta}$, and $\theta_{\kt}=\phi$.

\section{Numerical methods} \label{NUM}

We will calculate the reconstruction noise for sources at redshift: 1.25, 3 and 5. 
The power spectrum  covariance matrix will be measured from N-body simulations.
Unless mentioned explicitly, WMAP5 cosmological parameters are used throughout the paper:
$\Omega_m=0.258,\Omega_{\Lambda}=0.742,\Omega_b=0.0441,
\sigma_8=0.796,n_s=0.963,h=0.719,\tau=0.087$ \citep{2008arXiv0803.0586D}.

\subsection{Simulations} \label{SIMUL}

\begin{figure}
\begin{center}
\psfig{figure=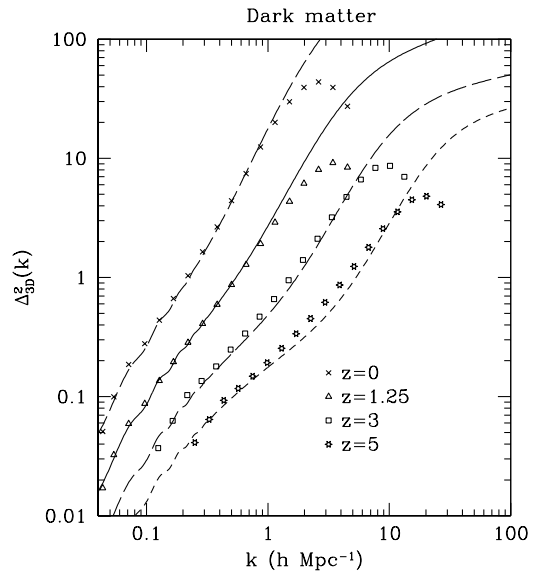,width=80mm,angle=0.}
\end{center}
\caption[Power spectrum at z=0-5]
{\footnotesize Power spectrum at z=0, 1.25, 3, and 5. The data points are from simulations, and the lines are 
the predicted non-linear power spectra generated by {\small CAMB}. The non-linearity increases
with lower redshift. The power spectra from simulations at small scales drop due to the resolution limits.
The parameters of the simulations are shown at table
\ref{tab:simulpar}. 

}
\label{fig:ps}
\end{figure}

We generate the N-body simulations with the {\small CUBEPM} code.  
{\small CUBEPM} is MPI paralleled 
particle-mesh (PM) code, and has particle-particle forces implemented at sub-grid scales.
It is further parallelized by shared-memory OpenMP on each node. 
The simulation volume (which is also called simulation box) is cubically 
decomposed to $n^3$ sub-volumes, and the calculation of each sub-volume is performed 
on one node of cluster. The total number of nodes used in simulation is $n^3$. 
The code has been run on up to 1000 nodes. 
We used $n=2$ and $3$ in our simulations. The 
simulations are run on the SUNNYVALE cluster of Canadian Institute for Theoretical Astrophysics (CITA), 
which is a Beowulf cluster composed of 200 Dell PE1950 nodes.  For each node there are 2 quad core Intel(R) Xeon(R) 
E5310 @ 1.60GHz processors, 4GB of RAM, and 2 gigE network interfaces.

HI gas is distributed in galaxies. If we ignore the bias 
between the galaxies and dark matter distributions, we could approximately use dark 
matter to represent 21-cm sources distributions at these
redshifts (the simple toy model in LP08) since HI sources 
are expected to trace dark matter fairly well. 
We have generated the source distribution with {\small CUBEPM} and we
output the 3-D particles distribution at redshifts 1.25, 3 and 5.  
To correctly measure the covariance matrix of power spectrum, we
generate $186$ independent simulations at $z=1.25$. 
Because the non-linearlity is less at higher redshift,
we have run $100$ simulations at $z=3$ and $90$ simulations at
$z=5$. 
The density field is produced by assigning the mass of particles 
to nearby grids with Cloud in Cell (CIC) method.

\begin{table*}
\begin{center}
\begin{tabular*}{1.00\textwidth} {@{\extracolsep{\fill}} |r|rrrr|rr|}  \hline \hline
redshift & $L(h^{-1}Mpc)$ & $\Delta^2(2\pi/L)$ &$n_{grid}$ & $\Delta^2(k_{Ny})$ & $k_s$ & $\Delta_{lin}^2(k_s)$ \\
\hline

0 & 300 & 0.012 & 512 & 219 &  0.14 & 0.46   \\
\hline

1.25 & 300& 0.0036 &512  &39.3 &  0.28 & 0.34  \\
\hline
3.0 & 100 & 0.015 & 1024 & 40.5 &  0.47 & 0.19 \\ 
\hline
5.0 & 50 &  0.020 & 1728  & 26.5 &  0.99 &  0.16 \\ 
\hline \hline
\end{tabular*}
\end{center}
\caption [Parameters of simulations]
{ \footnotesize Source simulation parameters. We choose these simulations to
  optimize the computation load while maintaining the required
  resolution. The simulation parameter choice is validated by looking
  at the convergence of  the Fisher information content of the power spectrum in both
  linear and non-linear scales. We found the upper limit for the
  non-linear power spectrum at the fundamental modes, which has the same
  size as the box, is around $0.02$; and the lower limit for non-linear
  power spectrum at the Nyquist frequency $k_{Ny}$, is around $20$.
  The values of non-linear power spectra are estimated from  {\small
    CAMB}. $k_s$ are the saturation scales of the Fisher information, as shown in Fig.
  \ref{fig:infor}. We also list the values of the linear power spectra at 
  these saturation scales for
  all four redshifts.}
\label{tab:simulpar}
\end{table*}

As we will illustrate later, the 3-D Fisher information (as defined in RH05) of the amplitude
of source power spectrum
has similarity with the reconstruction noise. We thus address the
convergence as a function of the number of simulations with a Fisher information plot. We check 
the values of $\sum_{k,k'\leq k_{max}}{\bf \bar C}^{-1}_{k,k'}$, where ${\bf \bar C}$ is the covariance matrix of the normalized power spectrum
$P(k)/\langle{P(k)}\rangle$, that is, the Fisher information is inversely
proportional to the variance of the amplitude of normalized  power spectrum. 
We have considered runs with different resolutions and box-sizes for each redshift.
The box-size needs to be big enough so that not much of the linear modes are cut-off by the limited box-size. 
On the other hand, the non-linear structure needs to be resolved at small enough scales so that the saturation
effect can be seen. We confirmed the convergence at linear scales by comparing the results with 
the Gaussian prediction,
and at non-linear scales by comparing with higher resolution
simulations. We found the upper limit for the non-linear power spectrum
at the fundamental modes, which has the same size as the box, is around
$\Delta^2\lesssim 0.02$; and the lower limit for non-linear power
spectrum at the Nyquist frequency $k_{Ny}$, is around $\Delta^2\gtrsim
20$. The simulations parameters we finally chose are given in table
\ref{tab:simulpar}, and  the power spectra are shown at 
Fig. \ref{fig:ps}. The values of the non-linear power spectrum are
estimated from  {\small CAMB}, and thus   
can be obtained before running simulations. 
The actual power spectra from N-body simulations are always lower than that at
$k_{Ny}$ because of the finite resolution. 
The sources at $z=1.25$ are produced with
simulations with $300h^{-1}$Mpc box, and $256^3$ particles on $512^3$
grids.  For $z=3$, because the matter distribution is more linear, we
need to increase the resolution to reach the non-linear structures 
at smaller scales.
We did this by using a smaller box
with $L=100h^{-1}$Mpc and more refined grids with $n_{grid}=1024$. At 
$z=5$, the non-linearity is lower, and we choose a $50 h^{-1}$Mpc box
with $1728^3$ grids. For the sources at $z=3$ and $z=5$, we have
assigned the particles distributions to $512^3$ grids though the
N-body simulations were performed on finer meshes for the sake of
computation efficiency. We also include the Fisher information at $z=0$,
using the same simulation parameters as those at $z=1.25$.

\begin{figure} 
\begin{center}
\psfig{figure=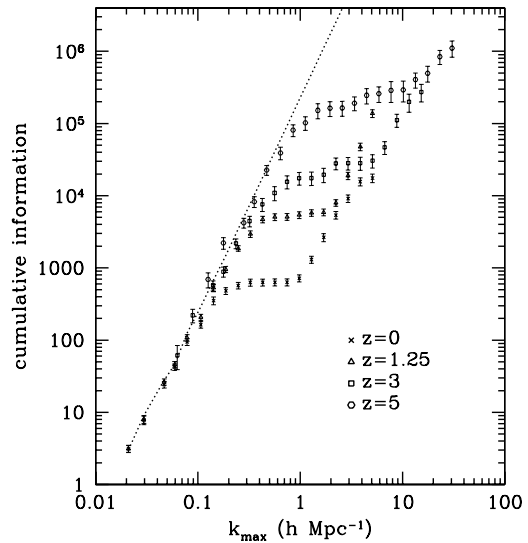,width=80mm,angle=0.}
\end{center}
\caption[3-D Fisher information content at z=0-5]
{\footnotesize The 3-D Fisher information content of the amplitude of power spectrum at $z=0, 1.25, 3.0, 5.0$ respectively.
All results are normalized to the values for a $300h^{-1}$Mpc box. The
dashed line is the Gaussian prediction. It is clear that the
Fisher information curves grow as $k^3$ at linear scales, then turn flat at
quasi-linear scales ($k\sim 0.5-2 h{\rm Mpc}^{-1}$ for $z=1.25$), before going up
again at non-linear scales. As expected, there is a gradual evolution of the plateau from $z=0$ to $z=5$.}
\label{fig:infor}
\end{figure}

The inversion of a large covariance matrix with a few hundred of
elements on each dimension can be numerically challenging. Following RH05, we 
divide the power spectra
to $N_b=20$ bins uniformly distributed in log scale, and calculate the
associated band power spectra and their covariance matrix. Subsequently, we invert the $N_b\times N_b$ matrix instead.
The Fisher information can be seen in Fig. \ref{fig:infor}. All results are normalized to the values for a $300h^{-1}$Mpc box, 
i.e., multiply the Fisher information measured by $(300/L)^3$ because larger
volume has more independent modes. It is clear that the Fisher information
curves grow as $k^3$ at linear scales, then turn flat at quasi-linear
scales ($k\sim 0.5-2 h{\rm Mpc}^{-1}$ for $z=1.25$), before going up again at
non-linear scales with a quasi-Gaussian scaling. There is a gradual evolution of the plateau from $z=0$ to $z=5$.

\begin{figure}
\begin{center}
\psfig{figure=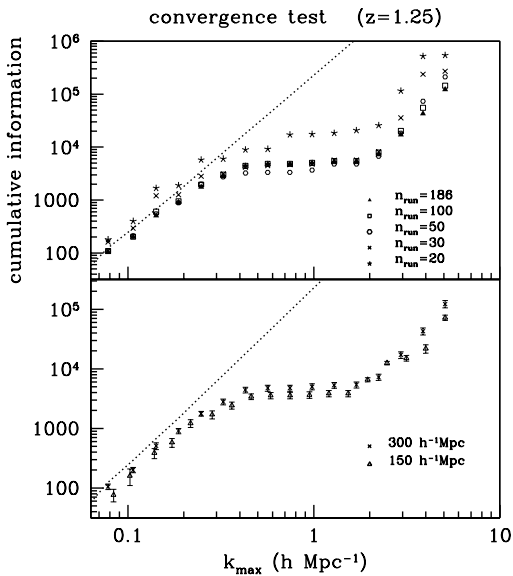,width=80mm,angle=0.}
\end{center}
\caption[Convergence tests of Fisher information content]
{\footnotesize Convergence test of the Fisher information as a function of simulations. We
  increase the number of simulations in the top panel. The relative 
  difference from the 186 runs result is 2\% for 100 runs, and 10\% for 50 runs.
  When the number of simulations is less than 30, the deviance
  is comparable to the Fisher information itself.  The bottom panel is for
  different box-sizes, both types have 186 runs. The error bars are generated by
  the bootstrap method. The Fisher information from
  simulations with $L=150h^{-1}$ Mpc agrees reasonably well with the
  one from $300h^{-1}$Mpc simulations. As such, the lower limit of
  power spectrum at fundamental mode should be equal to or higher than the value
  found in the $150h^{-1}$ Mpc simulations. }
\label{fig:infortest}
\end{figure}

We show the convergence of the Fisher information as a function of both the
number of simulations $n_{\rm sim}$ and the resolution in
Fig.~\ref{fig:infortest}. In the top panel, $n_{\rm sim}$ are reduced
from 186 to 100, 50, 30 and 20. The relative difference from the 186 runs result is 2\%
for 100 runs, and 10\% for 50 runs. When the number of simulations
is less than 30, the deviation is comparable to the Fisher information itself.    
We emphasize here that the required number of simulations also depends on the number of bins
$N_b$~: $N_b$ has to be much smaller than the number of simulations
for the covariance matrix to be non-singular. In the bottom panel, we
show the convergence of two sets of simulations with different
box-sizes. Both types have 186 runs.  The Fisher information from simulations
with $L=150h^{-1}$ Mpc agrees reasonably well with the one from
$300h^{-1}$Mpc simulations. We thus deduce that the lower limit for
the power spectrum at fundamental mode should be equal to or higher than the value found in
the $150h^{-1}$ Mpc simulations.  

\subsection{Lensing reconstruction} \label{RC}

We now study the lensing reconstruction noise that will be measured
directly from reconstructed lensing maps with the optimal quadratic
estimator described in section \ref{LENS}.  

\subsubsection{Power spectrum covariance matrix}

The  power spectrum covariance matrix can be written as
$\langle{P(\k)P(\k')}\rangle-\langle{P(\k)}\rangle\langle{P(\k')}\rangle$.\ Because
the source is isotropic and stationary, 
$\langle{P(\k)P(\k')}\rangle$ can be expressed in a
symmetric way ${\bf C}(k,k',\cos \theta_{\k,\k'})$, where $\cos
\theta_{\k,\k'}$ is the angle between the two vectors on the two shells
denoted by $k$ and $k'$.  We can then expand $\bf C$ in spherical harmonic
functions, and because it is independent of $\phi$, the expansion can
be written as the sum of the Legendre functions 
\beq
{\bf C}(k,k',\cos
\theta_{\k,\k'})=\sum_{l=0}^{\infty}{\bf C}_l(k,k')P_l(\cos\theta_{\k,\k'}).
\eeq 
Here $l$ is a even integer because ${\bf C}(k,k',\cos
\theta_{\k,\k'})$ is an even function of $\theta$. The window function can be written 
as an integral of the form
\beq
{\cal F}^{\kappa}(\k)  \propto \int {{\rm d}^3 k'\over (2\pi)^3}
 {\bf C}^{-1}(\k,\k') [G^{\kappa}_0(\k')+G^{\kappa}_2(\k') ]  \,,
\eeq 
where 
\beqa
G^{\kappa}_0(\k)& = & 2P+2P'k/3 \,, \\
G^{\kappa}_2(\k) & = & P'k(-\cos^2{\theta_{\k'}}+1/3) \,,
\eeqa
which are proportional to the zeroth order and second order Legendre
function $P_0(\cos\theta_{\k'})$ and $P_2(\cos\theta_{\k'})$. For each
$\k$, we can always choose the z-axis along it so that
$\cos\theta_{\k,\k'}$ is equal to $\cos\theta_{\k'}$ without loss of
generality. Because the orthogonal property of the Legendre functions, only
zeroth and second terms in ${\bf C}^{-1}$ remain, all the higher
orders cancel out by the integral over $\k'$.  

In other words, we decompose the optimal window function in two orthogonal components
from the zeroth and second order of the covariance matrix expansion terms respectively. 
For Gaussian sources, the power spectrum modes are uncorrelated with
other modes in different directions or on different shells, therefore all ${\bf C}_l$ are equal. 
For non-Gaussian sources, ${\bf C}_2$ is about an order of magnitude higher than ${\bf C}_0$, 
so the additional information of lensing obtained by using ${\bf C}_2$ is negligible 
(private communication with Joachim Harnois-Deraps, \citet{Jo2009}). 

\begin{figure} [t]
\begin{center}
\psfig{figure=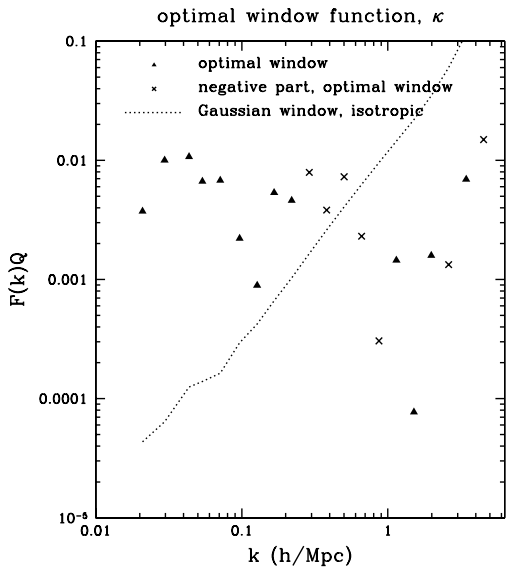,width=80mm,angle=0.}
\end{center}
\caption[Optimal window function in Fourier space]
{\footnotesize The optimal non-Gaussian 
window, and the optimal Gaussian window function in Fourier space.
The latter window has an almost power-law slope, while the former one 
has both positive and negative values due to the complicated behavior 
of inverse covariance matrix ${\bf C}^{-1}$. We use crosses to represent
the absolute values of the negative part of the optimal non-Gaussian window.
}
\label{fig:WinNG}
\end{figure}

In Fig. \ref{fig:WinNG}, we show the optimal non-Gaussian 
window, and the optimal Gaussian window function in Fourier space.
The latter window has an almost power-law slope, while the former one 
has both positive and negative values due to the complicated behavior 
of inverse covariance matrix ${\bf C}^{-1}$. We use crosses to represent
the absolute values of the negative part of the optimal non-Gaussian window.

In this paper, we will reconstruct the estimator using the zeroth
order only. Modes on a shell can be binned as a group first. We could
simplify the estimator by calculating its 1-D equivalent, e.g.,
replacing  $P(\k)$ with $P(k)$. Working in lower dimension facilitates
the numerical calculation, especially the inversion of covariance matrix. Rather
than working with a 6-D array ${\bf C}(\k,\k')$, we now only need to calculate the  covariance matrix
\beq
{\bf
  C}(k,k')=\langle{P(k)P(k')}\rangle-\langle{P(k)}\rangle\langle{P(k')}\rangle \,.
\eeq

We treat the modes on each $k$ shell as independent components and the window function corresponding to 
this shell is calculated from the overall contribution of these modes.
We use another trick  in the calculation of window function by
replacing $\sum {\bf C}^{-1}(\k,\k')G(\k')$ with $\sum {\bf C}_0^{-1}(k,k')G_0(k')$.

When considering the shear, the calculation is very similar. $G^{\gamma 1}=P'k\sin^2{\theta}\cos{2\phi}$
can also be written as the sum of Legendre functions: 
\beqa
G^{\gamma 1}_0 & = & 2P'k\cos{2\phi}/3 \,, \\
G^{\gamma 1}_2 & = & P'k(1/3-\cos^2{\theta})\cos{2\phi} \,.
\eeqa
Therefore only the zeroth and second order mode of covariance remain,
and we will also just use the zeroth order in the reconstruction  of
$\gamma_1$. For $\gamma_2$, one just need to replace $\cos{2\phi}$ by
$\sin{2\phi}$. The convergence can be calculated from the reconstructed
shear. We will call this convergence, $\gamma_E$. Similar to LP08, we
can choose the axis of the coordinate to be parallel to the direction of
the mode measured, so that $\gamma_E=\gamma_1$.   

\subsubsection{Convolution}

\begin{figure} 
\begin{center}
\psfig{figure=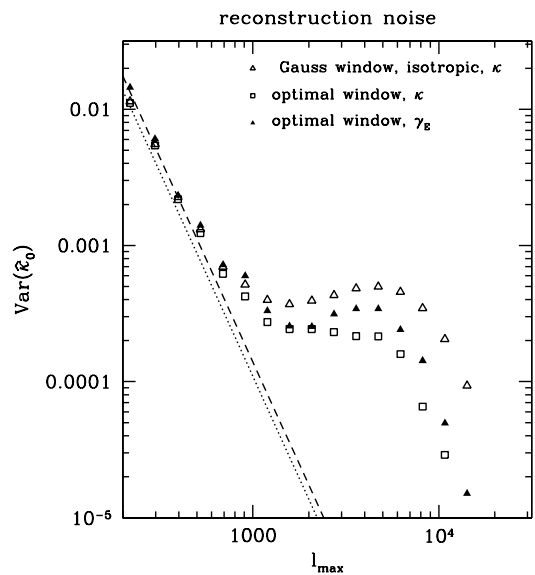,width=80mm,angle=0.}
\end{center}
\caption[Reconstruction noise of lensing at z=1.25]
{\footnotesize The contribution from one box at $z=1.25$ to the variance of
  various estimators. The variances are averaged over 186 independent
  boxes. The reconstruction noises for the Gaussian window, the optimal
  window for $\kappa$, and the optimal window function for
  $\gamma_{\rm E}$ are presented. At linear scales, i.e., when $\ell<1000$, the two
  estimators have the same reconstruction noise. The non-Gaussian
  window has a factor of a few less noise than the Gaussian estimator,
  and displays a plateau above the saturation scale. The plateau means
  that the non-Gaussianity of sources caused by the effective number
  of cells increases slowly below the saturation scales, i.e., the
  structures are intrinsically correlated below such scales and
  external factor such as experimental noise will not have much
  effect. The Gaussian estimator used in LP08 does not decrease at
  saturation scale, and even starts to increase at smaller scales. This
  is an artifact of the non-optimality of the Gaussian
  estimator for non-Gaussian sources. We also show the results from
  Gaussian sources as dashed (shear) and dotted lines (convergence).
  The slight non-monotonicity in the shear variance probably comes from
  the dropping of the quadrupolar $C_2$ term (see text).
}
\label{fig:noise_1.25}
\end{figure}

The optimal estimator contains the inverse  covariance matrix ${\bf
  C}^{-1}$, which we measure using around $100-200$ sources. For $512^3$ grids, there
are about 256 modes covering the range from the box-size scale to the
Nyquist scale. The inversion of the $256\times256$ covariance
matrix will be unstable, because the number of eigenvalues is more
than the total number of independent samples. As such, we will keep
using the 20 bins in our calculation. The binned window function are
less optimal, but the difference is a minor disadvantage compared
to the stability of the inversion. Another important term in the
window function is the gradient of the power spectra. Since the power
spectrum is close to a power-law, we calculate the gradient by finding
the tangent of the power spectrum at saturation scale $k_s$ measured
from the Fisher information on the  log-log plot. We find $P'k=n_{\rm NL}P$,
and $n_{\rm NL}=-1.6,-1.9$ and $-2.1$ for $z=1.25,3$ and $5$ respectively.

On the $512^3$ grids, we calculate the window function with the band
power, inverse matrix, and gradient. The Nearest Grid Point (NGP)
method is used  to map these band values to the 3-D grids in Fourier space.
From Eq.~(\ref{eq:angular}), we know that the optimal window function
does not only depend on the amplitude $k$, but is also a function of
$k_{\perp}$.  Note that when considering the shear, there is an extra angular
dependence on $\theta_{\kt}$, the direction on the transverse
plane that we need to take into account when computing the window  function.

Finally, the 3-D sources are convolved with the optimal window function. Since the window functions
are generated in Fourier space, we can simply transform the temperature fields to Fourier space by 
Fast Fourier Transform (FFT),  multiply by the window functions,
and transform them back to real space.  $\kappa$ or $\gamma_E$
maps are integrals of the covariance maps of the two convolved fields
along the line of sight. Since we are interested in the zero mode,
which is the average value of the covariance, we calculate the
reconstruction noise of the combined estimator of $\kappa$ and
$\gamma_E$ (as in LP08), which decrease the noise level. To validate
the procedure, we compare the reconstruction noises with the
analytical predictions in the Gaussian simulations with the same power
spectra as the simulated sources. 

\section{Numerical results and discussion} \label{DISC}

\subsection{Reconstruction noise} \label{NOISE}

The optimal estimators were derived in the constant $\kappa(\gamma)$
limit (zero-mode). For the other scales, we need to scale the
estimator by a normalization factor $b(\ell)$. The derivation of $b(\ell)$ 
can be found in appendix A. Not surprisingly, $b(\ell)\leq 1$,
and decreases from 1 to 0 when $\ell$ increases, for any given
estimator. For the scales we are interested in -- above the
characteristic scale $\ell_a$-- $b(\ell)$ is close to 1, and the noise
properties are similar to the zero-mode one. To illustrate the comparison between different estimators at various
$\ell$ with the of zero-mode reconstruction, we treat the experimental
noise as a hard cut-off at $k_c$. $\ell_a \sim \ell_c/2=\chi(z_s)k_c/2$. This
is motivated by the fact that the reconstructed noise is proportional
to $k^{-3}$ on linear scales, and the contribution to the
reconstruction at $k\leq k_c/2$ are small and can be
treated like the zero-mode. The values of $\ell_a$ are shown in Fig.~\ref{fig:cl_ngalaxy_z} and \ref{fig:dclratio}.

In Fig.~\ref{fig:noise_1.25}, the zero-mode $\kappa$ reconstruction noises
from various estimators are presented as functions of various cut-off
scales at $z=1.25$. For this plot, the noises correspond to the contribution
from a single simulation box of width $300h^{-1}\Mpc$, and is measured
over 186 independent boxes. As a reference, we present the
reconstruction noises from the  Gaussian window, non-Gaussian window
for $\kappa$ and the non-Gaussian window for $\gamma_{\rm E}$. On
linear scales, i.e., when $\ell<1000$, the two estimators of $\kappa$ have
the same reconstruction noise. At smaller scales, the non-Gaussian
window reduces the noise by factor of a few  compared to the
Gaussian estimator, and has a plateau after the saturation scale. The
experimental noise smears out the structure in the sources, and regions
which are originally independent become correlated. In other words,
lower noise level leads to a larger effective number of independent
source cells. Similarly, the effective number of cells increases if
the experiment is resolved at smaller scales, therefore  the
reconstructed noise is lower. The plateau means that the
non-Gaussianity of sources cause the effective number of cells to
increase very slowly below the saturation scale, i.e., the structures are
intrinsically correlated below this scale and external factors such as
experimental noise will have little effect. The Gaussian estimator we
used in LP08 also saturates at saturation scales, and even increases
at smaller scales. This is an artifact effect coming from
the fact that the Gaussian estimator is non-optimal for non-Gaussian
sources.  Note also that shear has better S/N level than $\kappa$.  
To compare with the simulated sources, we also show the results from
Gaussian predictions calculated using Wick's theorem. In fact, the
reconstruction noises in the Gaussian case can be approximated by the
following power-law relationship:  
\beqa
{\rm Var}(\kappa^{Gauss}_0) &\approx& 3 \pi^2 L^{-3} k_{c}^{-3} (1+{n_{\rm NL} \over 3})^{-2} \,,\nonumber \\
{\rm Var}(\gamma^{Gauss}_{E0}) &\approx& 45 \pi^2 L^{-3} k_{c}^{-3} n_{\rm NL} ^{-2} \,.
\eeqa 
Note that $n_{\rm NL}$ can not be -3 or 0, otherwise the variance of the estimator is
infinity and the lensing signal can not be reconstructed. An intuitive
explanation is that, in these cases the variance of temperature field are 
conserved even after being lensed, therefore lensing maps can not be distinguished
from unlensed ones and lensing can not be extracted.  
This stems from the fact that we only consider the zero order, ${\bf C}_0$, in the
covariance matrix of matter power spectrum. Lensing can still be
solved for $n_{\rm NL}=-3,0$ cases if ${\bf C}_2$ is taken into account in
the estimator.

\subsection{Saturation effects} \label{SATUR}

\begin{figure} [t]
\begin{center}
\psfig{figure=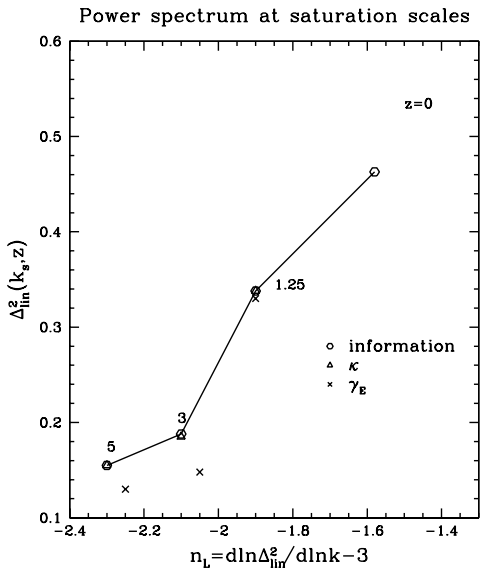,width=80mm,angle=0.}
\end{center}
\caption[Power spectrum at the saturation scale]
{\footnotesize Power spectra at the saturation scales of the lensing
  reconstruction noises for different redshifts. It can be seen that
  $\Delta^2_{lin}(k_s)$ increases with steeper shape of power
  spectrum. The points measured from the $\kappa$ saturation almost
  overlap with those measured from the Fisher information saturation because
  the shapes of the source power spectra are very close to power-laws.}
\label{fig:delta2ks}
\end{figure}

\begin{figure}
\begin{center}
\psfig{figure=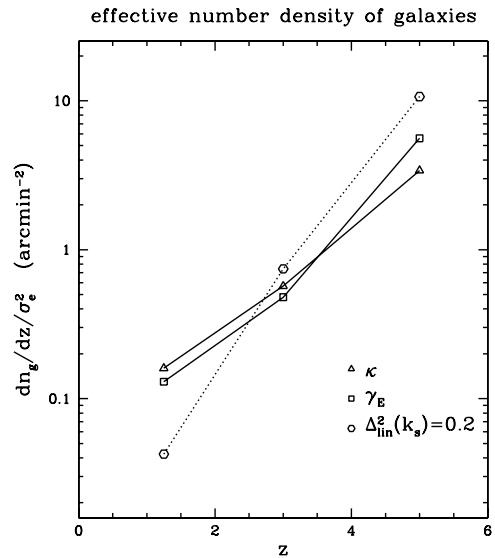,width=80mm,angle=0.}
\end{center}
\caption[Effective number density of sources versus redshift]
{\footnotesize The effective number densities of sources at various
  redshifts. When the redshift is higher, the non-linearity of sources goes 
  down. $dn_g/dz/\sigma^2_e$ increases quickly with redshift. We also
  plot the number density for a toy model where $\Delta^2_{lin}(k_s)=0.2$ for all redshifts. 
}
\label{fig:dngdz}
\end{figure}

We define the saturation scale $k_s$ as the wavenumber where the noise
from the Gaussian prediction is equal to the average amplitude of reconstructed noise in the
plateau. We have discussed the saturation effects of lensing 
reconstruction in LP08, where we found that at $z=7$ the S/N saturates at
the equivalent scale $k_s$ where the non-linear power spectrum of
source $\Delta^2(k_s)\sim 0.2$. The corresponding linear power
spectrum at $k_s$ is about $0.17$, which is consistent with the results
at  other redshifts shown in table \ref{tab:n_galaxy}. It is not clear
why the value  of $\Delta^2_{lin}(k_s)$ is slightly higher at z=7. It
could come from either the fact that there were less sources
($n_{sim}=20$) or that the estimator was less optimal in the works of LP08. 
Note that since we
used a Gaussian estimator in  LP08 and no ${\bf C}^{-1}$ term was
involved, the convergence of the reconstructed noise should be much
better than that for the Fisher information shown in Fig.~\ref{fig:infortest}. Since the
shape of the matter power spectrum persists except the non-linear
scales shift to larger scales, we expect to see similar saturation
effects at lower redshifts. This is confirmed by
Fig.~\ref{fig:noise_1.25} and later by Fig.~\ref{fig:delta2ks}.  

\begin{table*}
\begin{center}
\begin{tabular*}{1.00\textwidth} {@{\extracolsep{\fill}} |r||r|rrr|rr|}  \hline \hline
redshift & estimator &$d n_g/ dz/ {\sigma_e^2}  ({\rm arcmin}^{-2}) $ &$L_{cell} (h^{-1}{\rm Mpc})$ &$ \ell_s$ & $k_s$ &$\Delta^2_{lin}(k_s)$\\
\hline

1.25 & $\kappa$& 0.16  &  19.  & 773 & 0.28 &0.34 \\   
1.25 & $\gamma_E$& 0.13  &  20. & 773 & 0.28 & 0.34 \\ 
3.0 &$\kappa$& 0.57    & 13.  & 2136 &0.46 & 0.19 \\  
3.0 &$\gamma_E$& 0.48  & 14. & 1671 & 0.36 &  0.15 \\  
5.0 &$\kappa$ & 3.4  & 7.0 & 5657  & 0.99 &0.16 \\  
5.0 &$\gamma_E$ & 5.6 & 5.9  &   4514 & 0.79 & 0.13  \\ 
\hline
\hline
\end{tabular*}
\end {center}
\caption[Saturation scale and effective number density]
{\footnotesize Equivalent number densities 
of galaxies and saturation scales at different redshifts. 
The characteristic scale of the independent cell $L_{cell}$ is approximately  the saturation scale $k_s$.
The number density $n_g$ is calculated with Eq.~(\ref{eq:ng}). Because $n_g$ itself depends on the box-size of
the simulated source and the variance of intrinsic ellipticities $\sigma_e^2$, we compare $d n_g/ dz/ {\sigma_e^2}$
instead, which increases rapidly with redshift. 

}
\label{tab:n_galaxy}
\end{table*}
  
As we mentioned earlier, the saturation effect resembles the
Fisher information saturation effect for the initial amplitude of the 3-D
dark matter power spectrum in RH05.  Their cumulative Fisher information
increases at linear scales. On trans-linear scales, the Fisher information is
degenerate with that from larger scales and the cumulative Fisher information does
not increase. It is possible to view their Fisher information as a special
case of our lensing reconstruction calculation if the gradient term
$G$ equals the derivative of the power spectrum with regard to an
overall amplitude. \citet{2006MNRAS.370L..66N} have used halo models
to explain this saturation effect: on linear scales, the cumulative
Fisher information increases with higher $\ell$, as the volume and number
of halos do. On trans-linear scales, the 2-halo term first dominates,
which washes out the fluctuation in 1-halo term. On non-linear scales,
the contribution from small mass halos dominates, and the Fisher information
increases again $-$ at less than $1\%$ of the Gaussian information.  Our numerical results confirmed the reports of RH05. 

The saturation scales, as well as the non-linear scales, change with
redshifts. To illustrate this we investigate the evolution of
$\Delta^2_{lin}$ at the saturation scale. In Fig.~\ref{fig:delta2ks},
we plot $\Delta^2_{lin}(k_s)$ at three redshifts and we give the
corresponding numbers in table \ref{tab:n_galaxy}. Because the optimal
estimator is a function of the gradient of the power  spectrum, we
will try to see the evolution of $\Delta^2_{lin}(k_s)$ with the
gradient of power spectrum at the saturation scale.  It can be seen
that $\Delta^2_{lin}(k_s)$ increases with steeper shape of power
spectrum. The points from $\kappa$ almost overlap with those from
the Fisher information, because the shapes of source power spectra 
are very close to power-laws. 

\subsection{Effective number density} \label{NGALAXY}

\begin{figure}
\begin {center}
\psfig{figure=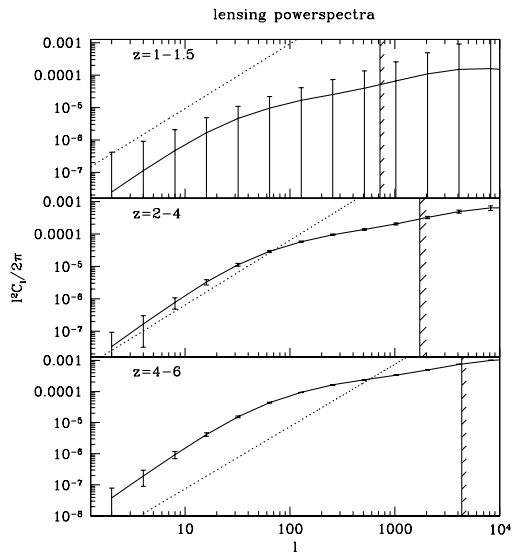,width=80mm,angle=0.}
\end {center}
\caption[Lensing signal and noise at redshift 1-5]
{\footnotesize The noise power spectra from the effective number densities of
  galaxies, compared with the lensing signals. The solid lines are the lensing
power spectra. The error bars are the contribution from both the lensing 
reconstruction noises and the cosmic variance. 
The noise dominates over the signal for sources at $z\sim 1-1.5$, becomes
comparable to but less than signal in $z\sim 2-4$, and further decreases at
$z\sim 4-6$. This corresponds to a sensitivity to lens structures around 
$z=0.5,1.0$ and $1.5$ respectively.
Note that the noise we display here is only valid for $\ell<\ell_a$, since
we calculate $n_g$ in the regime where the noise has similar behavior to zero mode.
$\ell_a$ are plotted as vertical lines here and in Fig. \ref{fig:dclratio}. 
}
\label{fig:cl_ngalaxy_z}
\end{figure}

\begin{figure} [t]
\begin {center}
\psfig{figure=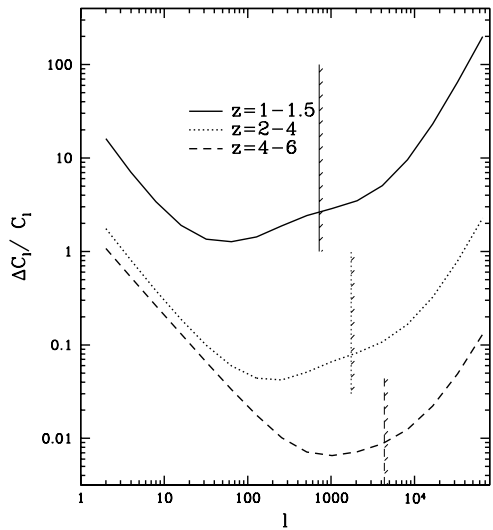,width=80mm,angle=0.}
\end {center}
\caption[The ratio of error bars and lensing signal]
{\footnotesize The ratio of the error bars and lensing signals. 
The vertical lines are $\ell_a$, and the noise model
is valid for $\ell<\ell_a$. For $\ell \gg \ell_a$, a scaling factor $b(\ell)$ need to be applied.
The calculation for $b(\ell)$ is done in LP08, and we found it
decreased from 1 gradually. Therefore the noises at these small scales
will be higher than what are shown in this plot. However, these scales
are not of interest to us in this paper. The error bars for $2-4$ bin
are at a few percents level for $\ell \sim 20-500$.  

}
\label{fig:dclratio}
\end{figure}

To describe the information content gained in the 21-cm lensing in a
more intuitive way, we define $n_g$, the equivalent effective surface
number density of galaxies which gives the same noise level at the
scale where S/N equals to one:  
\beq
{\rm Var}({\bar{ \hat \kappa}}) = \frac{\sigma_e^2}{n_g \cdot Area} \approx {1\over N_{cell}} \,.
\label{eq:ng}
\eeq
Similarly, we can define the number of effective independent cells, $N_{cell}$, and an effective number density
for the noise of $\gamma_E$. 
The characteristic scale of the
independent cell is approximately the saturation scale $k_s$. The cell size is 
\beqa
L_{cell} & = &2 \pi / k_s \times [27(n_{\rm NL}+3)^{-2} / 8 \pi]^{1/3} \,, \\
L_{cell}& = &2 \pi / k_s \times [45 n_{\rm NL}^{-2} / 8  \pi]^{1/3}\quad{\rm for}\quad \gamma_E \,,
\eeqa
where $\ell_s=k_s \chi(z)$ is the saturation scale that increases with
redshift.  $n_g$ at different redshifts are shown in table
\ref{tab:n_galaxy}. Because $n_g$ itself depends on the box-size of
the simulated source and the variance of the intrinsic ellipticities
$\sigma_e^2$, we compute $d n_g/ dz/ {\sigma_e^2}$ instead. From the
table, we see that the effective number density of galaxies
increases rapidly with redshift. 
We also plot
$dn_g/dz/\sigma^2_e$ versus $z$ in Fig.~\ref{fig:dngdz}, where one can see
that $dn_g/dz/\sigma^2_e$ 
increases quickly with redshift. We also show the number density for
a toy model where $\Delta^2_{lin}(k_s)=0.2$ for all redshifts.

In Fig.~\ref{fig:cl_ngalaxy_z}, we compare the shot noises $\ell^2 C_{\ell}^N/2
\pi$ from the effective number densities with the lensing signals. 
The error bars of the lensing power spectra are also plotted, which are estimated by
\beq
\Delta C_{\ell}= \sqrt{2 \over (2 \ell+1)\Delta \ell f_{sky}} (C_{\ell}^{\kappa}+C_{\ell}^N) \,,
\eeq
where $\Delta \ell\approx \ell/2$ because we used $\ell=2^n\,(n=1,2,...)$ bins, and 
consider a half sky survey 
\citep{2008PhRvL.100i1303C} with the fraction of the sky $f_{sky}=1/2$. 
Here we assume the noise of $\kappa$ is Gaussian, i.e., the eight-point function of temperature
is Gaussian though the temperature distribution itself is
non-Gaussian. Although not obvious, this point is supported by
numerically tests. We also neglect the non-Gaussianity of the
lensing observables themselves, which is not dominant at
$\ell<500$ but could become an issue at higher $\ell$ \citep{Dore2009}. It can
be seen that the noise dominates in $z\sim1-1.5$, becomes comparable to 
signal in $z\sim2-4$, and further decreases at $z\sim4-6$. For
redshift bin $z\sim 4-6$, we assume that ${\rm d}n_g/dz$ varies slowly, and use
$n_g=dn_g/dz|_{z=5} \Delta z$ with $\Delta z=2$, which should provide
the correct order of magnitude. Ideally, simulated sources should be
generated at all redshifts between $z=4$ and $z=6$, and the total number density is
then the integral of ${\rm d}n_g/dz$ over all z. The noise is calculated
similarly for $1-1.25$ and $2-4$ source redshift bins.

\begin{figure} [t]
\begin{center}
\psfig{figure=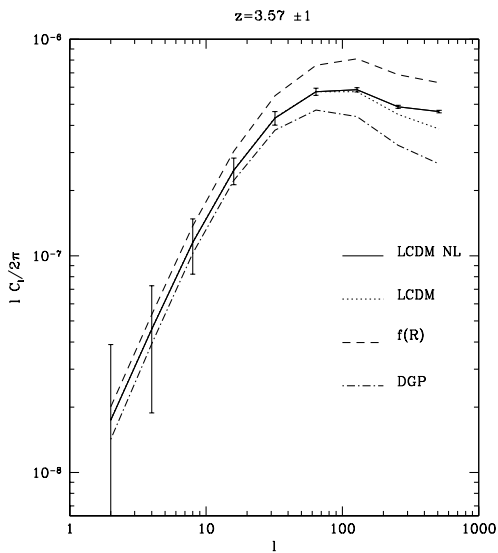,width=80mm,angle=0.}
\end{center}
\caption[Lensing power spectra from modified gravity models]
{\footnotesize The lensing power spectra from different models of dark energy (private communication
with Fabian Schmidt). The average redshift is 3.57, where the effective number density of sources is calculated by interpolation with the values from Fig. \ref{fig:dngdz}. The error bars are
 calculated in the
same way as  Fig. \ref{fig:cl_ngalaxy_z}.
}
\label{fig:cl_2-4}
\end{figure}

The noise is only valid for $\ell<\ell_a\sim \ell_s$, because we calculate
$n_g$ in the regime where the noise has a similar behavior to the zero mode.
We plot $\ell_a$ as vertical lines in Fig. \ref{fig:dclratio}. The full
noise calculation can be done similarly to LP08, however the zero-mode
$\kappa$  calculation works well enough because we are only interested
in large scales (small $\ell$'s). The noise at small scales will be due to
the increased non-Gaussianity at lower redshifts. For $\ell>\ell_a$, $b(\ell)$
is smaller than 1, and the noise will be higher. 

The relative error bars are shown in Fig. \ref{fig:dclratio}.  The
error bars for $z\sim 2-4$ bin are at a few percents level for $\ell \sim
20-500$. \citet{2008PhRvD..78d3002S} pointed out that linear scales
detection of lensing for source galaxies at $z\sim 1-3$ could be a good
way to distinguish three modified gravity models from a smooth dark
energy: $f(R)$ gravity, the DGP model, and the TeVeS theory. We show
the lensing power spectrum from different models of dark energy
(private communication with Fabian Schmidt) in
Fig. \ref{fig:cl_2-4}. The average redshift is 3.57, where the
effective number density of sources is calculated by interpolation
with the values 
from Fig. \ref{fig:dngdz}. The error bars are calculated in the same
way as  Fig. \ref{fig:cl_ngalaxy_z}. Because of the high precision of
the lensing reconstruction, we can expect to see a promising use of
21-cm sources to constrain modified gravity.

\begin{figure} [t]
\begin{center}
\psfig{figure=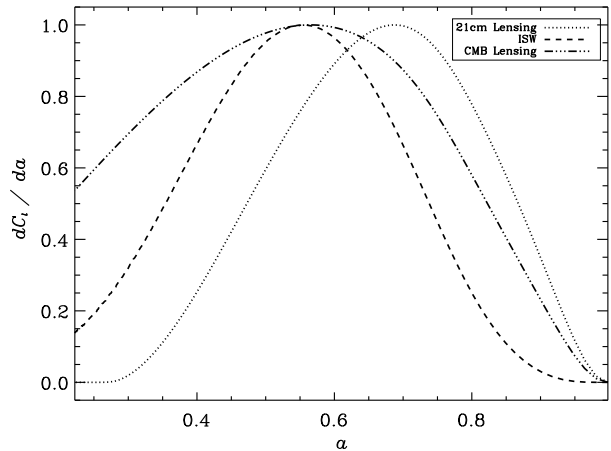,angle=90.,width=0.5\textwidth}
\end{center}
\caption[Lensing and ISW kernels]
{\footnotesize We plot the kernel for $\ell=20$ as defined in the text for the ISW,
  21-cm lensing ($2.5<z<3$) and the CMB lensing signals as a function
  of the scale factor. Whereas the CMB lensing overlapses with most of the
  ISW signal due to its width, we also see that the 21-cm lensing at
  high  $z$ has a very good overlapping with the ISW kernel which suggests
  that the 21-cm lensing from high $z$ constitutes a good probe of the
  ISW effect.}
\label{fig:kernel}
\end{figure}

Whereas we discussed here how to use 21-cm lensing to constrain dark energy from
measuring the lensing power spectrum. One can also investigate a parametrized
dark energy model by constraining $w_0,w_a$ using lensing tomography \citep{Dore2009}.
Another test of gravity and probe of dark-energy comes through the
Integrated Sachs-Wolfe (ISW) effect \citep{1967ApJ...147...73S}. In
Fig.~\ref{fig:kernel} we illustrate the potential for
cross-correlating the ISW signal using the lensing of sources with
$2.5<z<3$. As an illustration, we also plot the CMB lensing kernel. If
the $C_\ell$ of any of this signal is written as $C_\ell=\int da
W_{\ell}(a)$, we here plot $W_\ell(a)$ as a function of $a$ to illustrate the
sensitivity of our various probes. Whereas we see that the CMB lensing
kernel overlaps widely with the ISW signal, the 21-cm lensing signal is
a very good match to the ISW kernel. 
To quantify this, if we define as
the cross-correlation coefficient between two fields X and Y as
\beq
r^{XY}_\ell = \sum_\ell^{\ell_{max}}
C_\ell^{XY}/\sqrt{C_\ell^{XX}C_\ell^{YY}} \,,
\eeq
we find that the
cross-correlation coefficient at $\ell=20\, (40)$ for the CMB Lensing -
ISW and 21-cm lensing - ISW are respectively 0.98 (0.99) and 0.95
(0.96). As such, whereas neither are perfect, the 21-cm - ISW signal is
almost as good as the CMB - lensing. Using various redshift bins would
help us increasing this signal \citep{2004MNRAS.350.1445P}. 
By combining the different lensing sources planes, one can contruct an optimal
ISW estimator.  This is the subject of a paper in progress.

\section{Conclusion} \label{CONC}

In this paper, we discussed the possibility to constrain dark energy models  
with the lensing from 21-cm intensity emission at redshifts $z\sim 1-6$. First
we derived the optimal quadratic estimator for non-Gaussian sources,
which can be constructed numerically from simulations.  Then we
investigated the reconstruction noise with a large number of
simulations, and revealed that there is a saturation scale for the
reconstruction noise at all redshifts. We calculated the effective
number densities of 21-cm sources and compared the corresponding shot noises to the
lensing signals. We conclude that 21-cm sources
are promising lensing sources, because they can be observed at 
high redshift, and may soon be mapped over half the sky
\citep{2008PhRvL.100i1303C}. The lensing reconstructed from 21-cm sources at
$z\sim 3$, has a few percent error bars on linear scales at $z\sim 1$. 
These wide area dark matter maps are well suited to test dark energy dynamics
through cross correlation with the ISW effect, 
and can be
used to constrain modified gravity models. 

{\it Acknowledgments}
All computations were performed on the Canada Foundation for Innovation
funded CITA Sunnyvale cluster.
We thank Fabian Schmidt to provide the lensing power spectrum from modified gravity models in
Fig. 10, and Mark Neyrinck, Tzu-Ching Chang, Joachim Harnois-Deraps for helpful discussions.

\bibliography{lubib}

\begin{thebibliography}{25}
\expandafter\ifx\csname natexlab\endcsname\relax\def\natexlab#1{#1}\fi
\expandafter\ifx\csname bibnamefont\endcsname\relax
  \def\bibnamefont#1{#1}\fi
\expandafter\ifx\csname bibfnamefont\endcsname\relax
  \def\bibfnamefont#1{#1}\fi
\expandafter\ifx\csname citenamefont\endcsname\relax
  \def\citenamefont#1{#1}\fi
\expandafter\ifx\csname url\endcsname\relax
  \def\url#1{\texttt{#1}}\fi
\expandafter\ifx\csname urlprefix\endcsname\relax\def\urlprefix{URL }\fi
\providecommand{\bibinfo}[2]{#2}
\providecommand{\eprint}[2][]{\url{#2}}

\bibitem[{\citenamefont{{Bean} and {Dor{\'e}}}(2004)}]{2004PhRvD..69h3503B}
\bibinfo{author}{\bibfnamefont{R.}~\bibnamefont{{Bean}}} \bibnamefont{and}
  \bibinfo{author}{\bibfnamefont{O.}~\bibnamefont{{Dor{\'e}}}},
  \bibinfo{journal}{\prd} \textbf{\bibinfo{volume}{69}},
  \bibinfo{pages}{083503} (\bibinfo{year}{2004}),
  \eprint{arXiv:astro-ph/0307100}.

\bibitem[{\citenamefont{{Smith} et~al.}(2007)\citenamefont{{Smith}, {Zahn}, and
  {Dor{\'e}}}}]{2007PhRvD..76d3510S}
\bibinfo{author}{\bibfnamefont{K.~M.} \bibnamefont{{Smith}}},
  \bibinfo{author}{\bibfnamefont{O.}~\bibnamefont{{Zahn}}}, \bibnamefont{and}
  \bibinfo{author}{\bibfnamefont{O.}~\bibnamefont{{Dor{\'e}}}},
  \bibinfo{journal}{\prd} \textbf{\bibinfo{volume}{76}},
  \bibinfo{pages}{043510} (\bibinfo{year}{2007}), \eprint{0705.3980}.

\bibitem[{\citenamefont{{Pen}}(2004{\natexlab{a}})}]{2004NewA....9..417P}
\bibinfo{author}{\bibfnamefont{U.}~\bibnamefont{{Pen}}}, \bibinfo{journal}{New
  Astronomy} \textbf{\bibinfo{volume}{9}}, \bibinfo{pages}{417}
  (\bibinfo{year}{2004}{\natexlab{a}}).

\bibitem[{\citenamefont{{Hui} et~al.}(2007)\citenamefont{{Hui},
  {Gazta{\~n}aga}, and {Loverde}}}]{2007PhRvD..76j3502H}
\bibinfo{author}{\bibfnamefont{L.}~\bibnamefont{{Hui}}},
  \bibinfo{author}{\bibfnamefont{E.}~\bibnamefont{{Gazta{\~n}aga}}},
  \bibnamefont{and}
  \bibinfo{author}{\bibfnamefont{M.}~\bibnamefont{{Loverde}}},
  \bibinfo{journal}{\prd} \textbf{\bibinfo{volume}{76}},
  \bibinfo{pages}{103502} (\bibinfo{year}{2007}), \eprint{0706.1071}.

\bibitem[{\citenamefont{{Chang} et~al.}(2008)\citenamefont{{Chang}, {Pen},
  {Peterson}, and {McDonald}}}]{2008PhRvL.100i1303C}
\bibinfo{author}{\bibfnamefont{T.-C.} \bibnamefont{{Chang}}},
  \bibinfo{author}{\bibfnamefont{U.-L.} \bibnamefont{{Pen}}},
  \bibinfo{author}{\bibfnamefont{J.~B.} \bibnamefont{{Peterson}}},
  \bibnamefont{and}
  \bibinfo{author}{\bibfnamefont{P.}~\bibnamefont{{McDonald}}},
  \bibinfo{journal}{Physical Review Letters} \textbf{\bibinfo{volume}{100}},
  \bibinfo{pages}{091303} (\bibinfo{year}{2008}), \eprint{arXiv:0709.3672}.

\bibitem[{\citenamefont{{Wyithe} et~al.}(2008)\citenamefont{{Wyithe}, {Loeb},
  and {Geil}}}]{2008MNRAS.383.1195W}
\bibinfo{author}{\bibfnamefont{J.~S.~B.} \bibnamefont{{Wyithe}}},
  \bibinfo{author}{\bibfnamefont{A.}~\bibnamefont{{Loeb}}}, \bibnamefont{and}
  \bibinfo{author}{\bibfnamefont{P.~M.} \bibnamefont{{Geil}}},
  \bibinfo{journal}{\mnras} \textbf{\bibinfo{volume}{383}},
  \bibinfo{pages}{1195} (\bibinfo{year}{2008}).

\bibitem[{\citenamefont{{Tegmark} and
  {Zaldarriaga}}(2008)}]{2008arXiv0805.4414T}
\bibinfo{author}{\bibfnamefont{M.}~\bibnamefont{{Tegmark}}} \bibnamefont{and}
  \bibinfo{author}{\bibfnamefont{M.}~\bibnamefont{{Zaldarriaga}}},
  \bibinfo{journal}{ArXiv e-prints}  (\bibinfo{year}{2008}),
  \eprint{0805.4414}.

\bibitem[{\citenamefont{{Zhang} and {Pen}}(2005)}]{2005PhRvL..95x1302Z}
\bibinfo{author}{\bibfnamefont{P.}~\bibnamefont{{Zhang}}} \bibnamefont{and}
  \bibinfo{author}{\bibfnamefont{U.-L.} \bibnamefont{{Pen}}},
  \bibinfo{journal}{Physical Review Letters} \textbf{\bibinfo{volume}{95}},
  \bibinfo{pages}{241302} (\bibinfo{year}{2005}),
  \eprint{arXiv:astro-ph/0506740}.

\bibitem[{\citenamefont{{Cooray}}(2004)}]{2004NewA....9..173C}
\bibinfo{author}{\bibfnamefont{A.}~\bibnamefont{{Cooray}}},
  \bibinfo{journal}{New Astronomy} \textbf{\bibinfo{volume}{9}},
  \bibinfo{pages}{173} (\bibinfo{year}{2004}), \eprint{astro-ph/0309301}.

\bibitem[{\citenamefont{{Zahn} and {Zaldarriaga}}(2006)}]{2006ApJ...653..922Z}
\bibinfo{author}{\bibfnamefont{O.}~\bibnamefont{{Zahn}}} \bibnamefont{and}
  \bibinfo{author}{\bibfnamefont{M.}~\bibnamefont{{Zaldarriaga}}},
  \bibinfo{journal}{\apj} \textbf{\bibinfo{volume}{653}}, \bibinfo{pages}{922}
  (\bibinfo{year}{2006}), \eprint{arXiv:astro-ph/0511547}.

\bibitem[{\citenamefont{{Metcalf} and {White}}(2007)}]{2007MNRAS.381..447M}
\bibinfo{author}{\bibfnamefont{R.~B.} \bibnamefont{{Metcalf}}}
  \bibnamefont{and} \bibinfo{author}{\bibfnamefont{S.~D.~M.}
  \bibnamefont{{White}}}, \bibinfo{journal}{\mnras}
  \textbf{\bibinfo{volume}{381}}, \bibinfo{pages}{447} (\bibinfo{year}{2007}).

\bibitem[{\citenamefont{{Lu} and {Pen}}(2008)}]{2008MNRAS.388.1819L}
\bibinfo{author}{\bibfnamefont{T.}~\bibnamefont{{Lu}}} \bibnamefont{and}
  \bibinfo{author}{\bibfnamefont{U.-L.} \bibnamefont{{Pen}}},
  \bibinfo{journal}{\mnras} \textbf{\bibinfo{volume}{388}},
  \bibinfo{pages}{1819} (\bibinfo{year}{2008}), \eprint{arXiv:0710.1108}.

\bibitem[{\citenamefont{{Metcalf} and {White}}(2008)}]{2008arXiv0801.2571B}
\bibinfo{author}{\bibfnamefont{R.~B.} \bibnamefont{{Metcalf}}}
  \bibnamefont{and} \bibinfo{author}{\bibfnamefont{S.~D.~M.}
  \bibnamefont{{White}}}, \bibinfo{journal}{ArXiv e-prints}
  \textbf{\bibinfo{volume}{801}} (\bibinfo{year}{2008}), \eprint{0801.2571}.

\bibitem[{\citenamefont{{Rimes} and {Hamilton}}(2005)}]{2005MNRAS.360L..82R}
\bibinfo{author}{\bibfnamefont{C.~D.} \bibnamefont{{Rimes}}} \bibnamefont{and}
  \bibinfo{author}{\bibfnamefont{A.~J.~S.} \bibnamefont{{Hamilton}}},
  \bibinfo{journal}{\mnras} \textbf{\bibinfo{volume}{360}},
  \bibinfo{pages}{L82} (\bibinfo{year}{2005}), \eprint{arXiv:astro-ph/0502081}.

\bibitem[{\citenamefont{{Hui} et~al.}(2008)\citenamefont{{Hui},
  {Gazta{\~n}aga}, and {Loverde}}}]{2008PhRvD..77f3526H}
\bibinfo{author}{\bibfnamefont{L.}~\bibnamefont{{Hui}}},
  \bibinfo{author}{\bibfnamefont{E.}~\bibnamefont{{Gazta{\~n}aga}}},
  \bibnamefont{and}
  \bibinfo{author}{\bibfnamefont{M.}~\bibnamefont{{Loverde}}},
  \bibinfo{journal}{\prd} \textbf{\bibinfo{volume}{77}},
  \bibinfo{pages}{063526} (\bibinfo{year}{2008}), \eprint{0710.4191}.

\bibitem[{\citenamefont{{Loverde} et~al.}(2008)\citenamefont{{Loverde}, {Hui},
  and {Gazta{\~n}aga}}}]{2008PhRvD..77b3512L}
\bibinfo{author}{\bibfnamefont{M.}~\bibnamefont{{Loverde}}},
  \bibinfo{author}{\bibfnamefont{L.}~\bibnamefont{{Hui}}}, \bibnamefont{and}
  \bibinfo{author}{\bibfnamefont{E.}~\bibnamefont{{Gazta{\~n}aga}}},
  \bibinfo{journal}{\prd} \textbf{\bibinfo{volume}{77}},
  \bibinfo{pages}{023512} (\bibinfo{year}{2008}), \eprint{0708.0031}.

\bibitem[{\citenamefont{{Hu} and {Okamoto}}(2002)}]{2002ApJ...574..566H}
\bibinfo{author}{\bibfnamefont{W.}~\bibnamefont{{Hu}}} \bibnamefont{and}
  \bibinfo{author}{\bibfnamefont{T.}~\bibnamefont{{Okamoto}}},
  \bibinfo{journal}{\apj} \textbf{\bibinfo{volume}{574}}, \bibinfo{pages}{566}
  (\bibinfo{year}{2002}).

\bibitem[{\citenamefont{{Bertschinger}}(1992)}]{1992LNP...408...65B}
\bibinfo{author}{\bibfnamefont{E.}~\bibnamefont{{Bertschinger}}}, in
  \emph{\bibinfo{booktitle}{New Insights into the Universe}}, edited by
  \bibinfo{editor}{\bibfnamefont{V.~J.} \bibnamefont{{Martinez}}},
  \bibinfo{editor}{\bibfnamefont{M.}~\bibnamefont{{Portilla}}},
  \bibnamefont{and} \bibinfo{editor}{\bibfnamefont{D.}~\bibnamefont{{Saez}}}
  (\bibinfo{year}{1992}), vol. \bibinfo{volume}{408} of
  \emph{\bibinfo{series}{Lecture Notes in Physics, Berlin Springer Verlag}},
  pp. \bibinfo{pages}{65--+}.

\bibitem[{\citenamefont{{Dunkley} et~al.}(2008)\citenamefont{{Dunkley},
  {Komatsu}, {Nolta}, {Spergel}, {Larson}, {Hinshaw}, {Page}, {Bennett},
  {Gold}, {Jarosik} et~al.}}]{2008arXiv0803.0586D}
\bibinfo{author}{\bibfnamefont{J.}~\bibnamefont{{Dunkley}}},
  \bibinfo{author}{\bibfnamefont{E.}~\bibnamefont{{Komatsu}}},
  \bibinfo{author}{\bibfnamefont{M.~R.} \bibnamefont{{Nolta}}},
  \bibinfo{author}{\bibfnamefont{D.~N.} \bibnamefont{{Spergel}}},
  \bibinfo{author}{\bibfnamefont{D.}~\bibnamefont{{Larson}}},
  \bibinfo{author}{\bibfnamefont{G.}~\bibnamefont{{Hinshaw}}},
  \bibinfo{author}{\bibfnamefont{L.}~\bibnamefont{{Page}}},
  \bibinfo{author}{\bibfnamefont{C.~L.} \bibnamefont{{Bennett}}},
  \bibinfo{author}{\bibfnamefont{B.}~\bibnamefont{{Gold}}},
  \bibinfo{author}{\bibfnamefont{N.}~\bibnamefont{{Jarosik}}},
  \bibnamefont{et~al.}, \bibinfo{journal}{ArXiv e-prints}
  (\bibinfo{year}{2008}), \eprint{0803.0586}.

\bibitem[{\citenamefont{{Harnois-Deraps}
  et~al.}(2009)\citenamefont{{Harnois-Deraps}, {Pen}, and {Dore}}}]{Jo2009}
\bibinfo{author}{\bibfnamefont{J.}~\bibnamefont{{Harnois-Deraps}}},
  \bibinfo{author}{\bibfnamefont{U.-L.} \bibnamefont{{Pen}}}, \bibnamefont{and}
  \bibinfo{author}{\bibfnamefont{O.}~\bibnamefont{{Dore}}},
  \bibinfo{journal}{in preparation}  (\bibinfo{year}{2009}).

\bibitem[{\citenamefont{{Neyrinck} et~al.}(2006)\citenamefont{{Neyrinck},
  {Szapudi}, and {Rimes}}}]{2006MNRAS.370L..66N}
\bibinfo{author}{\bibfnamefont{M.~C.} \bibnamefont{{Neyrinck}}},
  \bibinfo{author}{\bibfnamefont{I.}~\bibnamefont{{Szapudi}}},
  \bibnamefont{and} \bibinfo{author}{\bibfnamefont{C.~D.}
  \bibnamefont{{Rimes}}}, \bibinfo{journal}{\mnras}
  \textbf{\bibinfo{volume}{370}}, \bibinfo{pages}{L66} (\bibinfo{year}{2006}),
  \eprint{arXiv:astro-ph/0604282}.

\bibitem[{\citenamefont{{Dore} et~al.}(2009)\citenamefont{{Dore}, {Lu}, and
  {Pen}}}]{Dore2009}
\bibinfo{author}{\bibfnamefont{O.}~\bibnamefont{{Dore}}},
  \bibinfo{author}{\bibfnamefont{T.}~\bibnamefont{{Lu}}}, \bibnamefont{and}
  \bibinfo{author}{\bibfnamefont{U.-L.} \bibnamefont{{Pen}}},
  \bibinfo{journal}{0905.0501}  (\bibinfo{year}{2009}).

\bibitem[{\citenamefont{{Schmidt}}(2008)}]{2008PhRvD..78d3002S}
\bibinfo{author}{\bibfnamefont{F.}~\bibnamefont{{Schmidt}}},
  \bibinfo{journal}{\prd} \textbf{\bibinfo{volume}{78}},
  \bibinfo{pages}{043002} (\bibinfo{year}{2008}), \eprint{0805.4812}.

\bibitem[{\citenamefont{{Sachs} and {Wolfe}}(1967)}]{1967ApJ...147...73S}
\bibinfo{author}{\bibfnamefont{R.~K.} \bibnamefont{{Sachs}}} \bibnamefont{and}
  \bibinfo{author}{\bibfnamefont{A.~M.} \bibnamefont{{Wolfe}}},
  \bibinfo{journal}{\apj} \textbf{\bibinfo{volume}{147}}, \bibinfo{pages}{73}
  (\bibinfo{year}{1967}).

\bibitem[{\citenamefont{{Pen}}(2004{\natexlab{b}})}]{2004MNRAS.350.1445P}
\bibinfo{author}{\bibfnamefont{U.-L.} \bibnamefont{{Pen}}},
  \bibinfo{journal}{\mnras} \textbf{\bibinfo{volume}{350}},
  \bibinfo{pages}{1445} (\bibinfo{year}{2004}{\natexlab{b}}),
  \eprint{arXiv:astro-ph/0402008}.

\end{thebibliography}


\label{lastpage}

\end{document}